\documentclass[3p,twocolumn,number,sort&compress]{elsarticle}
\usepackage{epsfig}
\usepackage{subfigure}
\usepackage{amsmath}
\usepackage{color}
\usepackage{verbatim}
\usepackage{amssymb}
\usepackage{setspace}
\usepackage{graphicx}
\usepackage{subfigure}
\usepackage{dcolumn}
\usepackage{bm}
\usepackage{times}
\usepackage{enumerate}
\input epsf

\graphicspath{{figures/}}

\begin{document}

\author[Trinity,URV]{Per Sebastian Skardal}
\ead{skardals@gmail.com} 
\author[UNC]{Dane Taylor}
\author[Clarkson]{Jie Sun} 
\author[URV]{Alex Arenas} 
\address[Trinity]{Department of Mathematics, Trinity College, Hartford, CT 06106, USA}
\address[URV]{Departament d'Enginyeria Informatica i Matem\'{a}tiques, Universitat Rovira i Virgili, 43007 Tarragona, Spain}
\address[UNC]{Department of Mathematics, University of North Carolina, Chapel Hill, NC 27599, USA}
\address[Clarkson]{Department of Mathematics, Clarkson University, Potsdam, NY 13699, USA}

\title{Erosion of synchronization: Coupling heterogeneity and network structure}


\begin{abstract}
We study the dynamics of network-coupled phase oscillators in the presence of coupling frustration. It was recently demonstrated that in heterogeneous network topologies, the presence of coupling frustration causes perfect phase synchronization to become unattainable even in the limit of infinite coupling strength. Here, we consider the important case of heterogeneous coupling functions and extend previous results by deriving analytical predictions for the total erosion of synchronization. Our analytical results are given in terms of basic quantities related to the network structure and coupling frustration. In addition to fully heterogeneous coupling, where each individual interaction is allowed to be distinct, we also consider partially heterogeneous coupling and homogeneous coupling in which the coupling functions are either unique to each oscillator or identical for all network interactions, respectively. We demonstrate the validity of our theory with numerical simulations of multiple network models, and highlight the interesting effects that various coupling choices and network models have on the total erosion of synchronization. Finally, we consider some special network structures with well-known spectral properties, which allows us to derive further analytical results.
\end{abstract}


\maketitle

\section{Introduction}\label{sec:1}
Self-organization and emergent collective behavior represent universal concepts that are vital in many nonlinear processes~\cite{Strogatz2003,Pikovsky2001}. Synchronization of large ensembles of coupled oscillators plays a particularly important role in our understanding of complex and network-coupled dynamical systems~\cite{Arenas2008PR}. Examples of the importance of synchronization can be found in natural phenomena, for instance the functionality of cardiac pacemakers~\cite{Glass1988}, mammalian circadian rhythms~\cite{Yamaguchi2003Science}, and rhythmic flashing of fireflies~\cite{Buck2988QRB}, as well as engineered systems, for instance arrays of Josephson junctions~\cite{Wiesenfeld1996PRL}, the power grid~\cite{Motter2013NaturePhysics}, and pedestrian bridges~\cite{Strogatz2005Nature}. A particularly useful model for studying the synchronization of nonidentical oscillators was developed by Kuramoto~\cite{Kuramoto1984}, who showed that under suitable conditions the dynamics of $N$ coupled oscillators can be reduced to the dynamics of $N$ phase angles $\theta_i$, for $i=1,\dots,N$. When placed on a network whose structure dictates the oscillators' interaction patterns, the evolution of each phase is given by
\begin{align}
\dot{\theta}_i = \omega_i + K\sum_{j=1}^NA_{ij}H_{ij}(\theta_j-\theta_i),\label{eq:Kuramoto}
\end{align}
where the natural frequency $\omega_i$ describes the prefered frequency of oscillator $i$ in the absence of coupling, $K\ge0$ is the global coupling strength, the adjacency matrix $A_{ij}$ encodes the network interactions, which is assumed to be undirected such that $A_{ij}=A_{ji}$, and $H_{ij}(\theta)$ is the coupling function that describes the functional effect of oscillator $j$ on oscillator $i$, which is assumed to be $2\pi$-periodic and continuously differentiable.

The dynamics exhibited by Eq.~(\ref{eq:Kuramoto}) have been studied in various contexts~\cite{Moreno2004EPL,Ichinomiya2004PRE,Restrepo2005PRE,Oh2005PRE,Arenas2006PRL,GomezGardenes2007PRL,GomezGardenes2011PRL,Barlev2011Chaos,Rohden2012PRL,Skardal2012PRE,Witthaut2012NJP,Nicosia2013PRL,Skardal2013EPL,Dorfler2013PNAS,Jorg2014PRL,Skardal2014PRL,Skardal2014PRE,Restrepo2014EPL,Skardal2015Arxiv} and have advanced our understanding of collective behavior, particularly regarding the interplay between structure and dynamics and their effects on synchronization. Typically, the extent of phase synchronization of the oscillators is measured by the classical Kuramoto order parameter $r$ that is defined by~\cite{Kuramoto1984}
\begin{align}
re^{i\psi} = \frac{1}{N}\sum_{j=1}^Ne^{i\theta_j},\label{eq:OrderParameter}
\end{align}
where the complex number $re^{i\psi}$ represents the oscillators' centroid in the complex unit circle. In particular, the order parameter $r$ ranges from $0$ to $1$, indicating complete incoherence and perfect synchronization, respectively, while intermediate values typically correspond to partial synchronization. Alternatively, several studies have defined the degree of phase synchronization using a combination of the collection of local order parameters, defined $r_ie^{i\psi_i}=\sum_{j=1}^NA_{ij}e^{i\theta_j}$ for $i=1,\dots,N$~\cite{Restrepo2005PRE,Skardal2013EPL,Skardal2014PRE}.

A key element of the model in Eq.~(\ref{eq:Kuramoto}) is the choice of coupling functions $H_{ij}(\theta)$ that defines the interactions between oscillators. For instance, the choice $H_{ij}(\theta)=\sin(\theta)$ yields the classical Kuramoto model~\cite{Kuramoto1984}, while the presence of additional modes can give rise to multi-branch entrainment, a.k.a. cluster synchronization~\cite{Daido1996PRL,Aonishi2001PRL,Skardal2011PRE,Laing2012Chaos,Komarov2013PRL,Lai2013PRE,Li2014PRE,Komarov2014PhysicaD}. Here, we focus our attention on systems with coupling frustration, as indicated by one or more non-zero values of the quantity
\begin{equation}
h_{ij}=H_{ij}(0)/\sqrt{2}H_{ij}'(0). 
\end{equation}
The physical interpretation of coupling frustration corresponds to the case where the networks' interaction terms do not all vanish when all phases are equal. The presence of coupling frustration is vital in the modeling of excitable and reaction-diffusion dynamics for the reason that neighboring elements typically do not react simultaneously, but rather one after another~\cite{Kopell2002}. Many such examples exist in biological and chemical systems, including neuron excitation~\cite{FitzHugh1955}, cardiac dynamics~\cite{Karma2007}, and the Belousov-Zhabotinsky reaction~\cite{Winfree1984}. Additionally, coupling frustration has been linked to the emergence of chimera states~\cite{Shima2004PRE,Abrams2004PRL,Abrams2008PRL,Laing2009,Martens2010PRL,Panaggio2013PRL,Panaggio2015PRL,Buscarino2015PRE,Xie2015PRE}, non-universal synchronization transitions~\cite{Omelchecnko2012PRL}, and other effects~\cite{Vlasov2014Chaos}.

In a recent publication~\cite{Skardal2015PRE} we reported a novel phenomenon for networks of coupled oscillators that we called {\it erosion of synchronization}. In particular, we found that in the presence of both coupling frustration and structural heterogeneity the perfectly synchronized state (i.e., $r=1$, or equivalently, $\theta_1=\theta_2=\dots=\theta_N$) becomes unattainable in steady-state even in the limit of infinite coupling strength. To quantify the total erosion of synchronization in a network, we consider the quantity $1-r$ in the limit $K\to\infty$, denoted $1-r^\infty$. We demonstrated this by considering the case of \emph{homogeneous coupling}, i.e., $H_{ij}(\theta)=H(\theta)$, and subsequently showed that the total erosion of synchronization could be separated into the product of two terms describing the contributions of coupling frustration and structural heterogeneity, respectively, and that both of these terms amplify the total erosion of synchronization. 

In this Article, we provide a more complete description of this phenomenon. In particular, we extend our previous results to account for the important case of heterogeneous coupling, i.e., when the coupling function governing the interaction between each pair of network neighbors may be distinct. We refer to this most general case, where each $H_{ij}(\theta)$ is potentially different, as {\it full coupling heterogeneity}. In this case we assume that each undirected link has an associated coupling function, so that $H_{ij}(\theta)=H_{ji}(\theta)$. We also treat the case where each oscillator has its own coupling function, i.e., $H_{ij}(\theta)=H_i(\theta)$, which we refer to as {\it partial coupling heterogeneity}. Unlike the homogeneous coupling case, in both the fully and partially heterogeneous coupling cases we find that the total erosion of synchronization cannot be separated into a product of contributions from the coupling frustration and structural heterogeneity.

The remainder of this Article is organized as follows. In Section~\ref{sec:2} we present our theoretical results, which extend previous results for homogeneous coupling to the cases of both full and partial coupling heterogeneity. In Section~\ref{sec:3} we present results from numerical simulations that support our theory and explore the interplay between coupling frustration and structural heterogeneity. In Section~\ref{sec:4} we study the stability of the synchronized state. In Section~\ref{sec:5} we investigate erosion of synchronization in several network models with well-known spectral properties, allowing us to develop further analytical results. In particular, we consider the star and chain networks, as well as Watts-Strogatz networks~\cite{Watts1998Nature}. Finally, in Section~\ref{sec:6} we conclude with a discussion of our results.

\section{Theory}\label{sec:2}
In this Section we present a theoretical framework for quantifying the erosion of synchronization for the dynamics defined in Eq.~(\ref{eq:Kuramoto}). We begin by considering the case of fully heterogeneous coupling, i.e., where each undirected link connecting oscillators $i$ and $j$ have a potentially distinct coupling function $H_{ij}(\theta)=H_{ji}(\theta)$. We also consider the case of partially heterogeneous coupling, i.e., when each oscillator has its own coupling functions, $H_{ij}(\theta)=H_i(\theta)$. Finally, we compare these results to the originally derived results for homogeneous coupling, i.e., $H_{ij}(\theta)=H(\theta)$, presented in Ref.~\cite{Skardal2015PRE}.

\subsection{Fully heterogeneous coupling}\label{subsec:2.1}
We begin by following Ref.~\cite{Skardal2014PRL} and consider the dynamics of Eq.~(\ref{eq:Kuramoto}) in the strong coupling regime, i.e., $r\approx1$. In typical networks, such a state can be attained in a variety of ways, most readily by considering either a sufficiently large coupling strength, or a set of natural frequencies with a sufficiently small spread. It is worth pointing out that these two situations are equivalent up to a rescaling of time, and thus the results presented here are valid in both cases. In the strong coupling regime the oscillators become tightly packed around the mean phase $\psi$, implying that $|\theta_i-\theta_j|\ll1$ for all $(i,j)$ pairs. Thus, the contribution of each pair-wise interaction can be linearized to $H_{ij}(\theta_j-\theta_i)\approx H_{ij}(0) + H_{ij}'(0)(\theta_j-\theta_i)$, and Eq.~(\ref{eq:Kuramoto}) can be approximated by
\begin{align}
\dot{\theta}_i \approx \omega_i+ K\tilde{d}_i - K\sum_{j=1}^N\tilde{L}_{ij}\theta_j,\label{eq:Theory01}
\end{align}
or rather in vector form,
\begin{align}
\dot{\bm{\theta}}\approx\bm{\omega}+K\bm{\tilde{d}}-K\tilde{L}\bm{\theta}.\label{eq:Theory02}
\end{align}
Here,  $\bm{\tilde{d}}$ and $\tilde{L}$ represent the \textit{weighted degree vector} and \textit{weighted Laplacian matrix}. In contrast to the unweighted degree vector $\bm{d}$ and unweight Laplacian matrix $L$, whose entries are defined
\begin{align}
d_i=\sum_{j=1}^N A_{ij},\hskip4ex L_{ij}=\delta_{ij}d_i-A_{ij},\label{eq:Theory03}
\end{align}
the entries of their weighted counterparts are given by
\begin{gather}
\tilde{d}_i=\sum_{j=1}^NA_{ij}H_{ij}(0),\label{eq:Theory04}\\
\tilde{L}_{ij}=\delta_{ij}\left(\sum_{l=1}^NA_{il}H_{il}'(0)\right) - A_{ij}H_{ij}'(0),\label{eq:Theory05}
\end{gather}
where $\delta_{ij}$ is the Kronecker delta.

We now aim to solve for the steady-state of Eq.~(\ref{eq:Theory02}) where, in a synchronized state, each oscillator travels at the same effective frequency, $\dot{\theta}_i=\Omega$ for all $i=1,\dots,N$. This effective frequency, which is also given by $\Omega=\langle\dot{\theta}\rangle=N^{-1}\sum_{i}\dot{\theta}_i$ is simply the mean of the heterogeneous part of Eq.~(\ref{eq:Theory02}), i.e., $\Omega=\langle\omega+Kd\rangle=N^{-1}\sum_{i=1}^N\omega_i+Kd_i$, which follows from the fact that $\tilde{L}$ is symmetric and maps vectors to the space of vectors with zero mean (see explanation below). Inserting $\dot{\bm{\theta}}=\bm{\Omega}=\Omega[1,\dots,1]^T$ in Eq.~(\ref{eq:Theory02}) and rearranging, we effectively enter the rotating reference frame $\bm{\theta}\mapsto\bm{\theta}+\bm{\Omega}t$ and obtain
\begin{align}
\tilde{L}\bm{\theta}^*=(\bm{\omega}+K\bm{\tilde{d}}-\bm{\Omega})/K,\label{eq:Theory06}
\end{align}
where $\bm{\theta}^*$ denotes the steady-state solution in the rotating frame.

To solve for $\bm{\theta}^*$, we now aim to define the pseudoinverse of the weighted Laplacian matrix~\cite{BenIsrael1974}, which can be defined using the spectral properties of $\tilde{L}$. Assuming that the network is connected and undirected, $\tilde{L}$ has a single zero eigenvalue with the remainder being real and positive so that they can be ordered $0=\lambda_1<\lambda_2\le\dots\le\lambda_N$. Furthermore, the eigenvector that corresponds to the trivial eigenvalue $\lambda_1=0$ is simply the constant vector $\bm{v}^1\propto[1,\dots,1]^T$ and represents motion along the synchronization manifold. With the remainder of the eigenvectors (which we assume are normalized to $\|\bm{v}^j\|=1$) the pseudoinverse can be defined
\begin{align}
\tilde{L}^\dagger=\sum_{j=2}^N\frac{\bm{v}^j\bm{v}^{jT}}{\lambda_j}.\label{eq:Theory07}
\end{align}
Importantly, we note that $\tilde{L}$ and $\tilde{L}^\dagger$ share the nullspace of the all vectors spanned by $\bm{v}^1$, and thus map all vectors to the $N-1$-dimensional space of vectors in $\bm{R}^N$ with zero mean.

We can now return to Eq.~(\ref{eq:Theory06}) and apply the pseudoinverse to obtain
\begin{align}
\bm{\theta}^*=\tilde{L}^\dagger\bm{y},\label{eq:Theory10}
\end{align}
where
\begin{align}
\bm{y}=\bm{\omega}/K + \bm{\tilde{d}}.\label{eq:Theory11}
\end{align}
We note that $\bm{\Omega}$ does not appear in Eq.~(\ref{eq:Theory11}) since $\tilde{L}^\dagger\bm{\Omega}=\bm{0}$. To evaluate the degree of synchronization, we now consider the order parameter given by Eq.~(\ref{eq:OrderParameter}). We note that by a suitable shift in initial conditions, the mean phase $\psi$ can be set to zero without loss of generality. Noting that both sides of Eq.~(\ref{eq:OrderParameter}) must be purely real, the right-hand side of Eq.~(\ref{eq:OrderParameter}) can be expanded to obtain
\begin{align}
r\simeq1-\frac{\|\bm{\theta}^*\|^2}{2N}.\label{eq:Theory12}
\end{align}
Inserting Eq.~(\ref{eq:Theory02}) into Eq.~(\ref{eq:Theory03}), noting that $\|\bm{\theta}^*\|^2=\langle\bm{\theta}^*,\bm{\theta}^*\rangle$, and using the definition of $\tilde{L}^\dagger$ in Eq.~(\ref{eq:Theory07}), we obtain
\begin{align}
r\simeq1-\frac{J(\bm{y},\tilde{L})}{2},\label{eq:Theory13}
\end{align}
where $J$ is the {\it synchrony alignment function} previously studied in Refs.~\cite{Skardal2014PRL,Skardal2015PRE} and is defined by
\begin{align}
J(\bm{y},\tilde{L}) = \frac{1}{N}\sum_{j=2}^N\frac{\langle\bm{v}^j,\bm{y}\rangle^2}{\lambda_j^2}.\label{eq:Theory14}
\end{align}
Finally, to evaluate the total erosion of synchronization we consider the deviation from perfect synchronization, $1-r$, in the limit $K\to\infty$, which we denote $1-r^\infty$. In this limit we have that $\bm{y}=\bm{\tilde{d}}$, and we finally obtain
\begin{align}
1-r^\infty \simeq \frac{J(\bm{\tilde{d}},\tilde{L})}{2}.\label{eq:Theory16}
\end{align}

Equation~(\ref{eq:Theory16}) quantifies the total erosion of synchronization in a given network. We point out that the contributions from the coupling frustration [i.e., $H_{ij}(0)$ and $H_{ij}'(0)$] and the network structure (i.e., $\bm{d}$ and $L$) do not separate, as they jointly define the entries of both the weighted degree vector $\bm{\tilde{d}}$ and the weighted Laplacian $\tilde{L}$.

\begin{figure*}[t]
\centering
\includegraphics[width=0.85\linewidth]{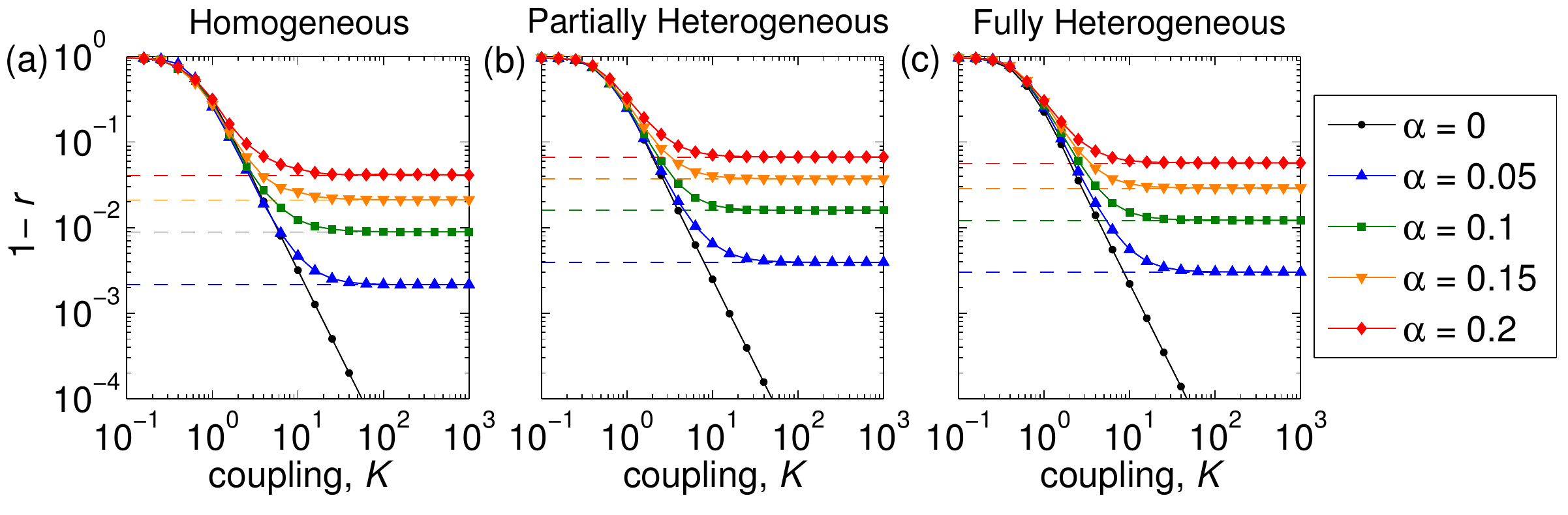}
\caption{(Color online) \textit{Erosion of synchronization: Effect of coupling heterogeneity and frustration.} Deviation from perfect synchronization $1-r$ vs coupling strength $K$ in a SF network of size $N=1000$, exponent $\gamma=3$, and mean degree $\langle d\rangle=4$. Results are presented for (a)~homogeneous coupling, (b) partially heterogeneous coupling, and (c) fully heterogeneous coupling for phase-lag parameter $\alpha=0$ (black circles), $0.05$ (blue triangles), $0.1$ (green squares), $0.15$ (orange inverted triangles), and $0.2$ (red diamonds). Results for partially and fully heterogeneous couplings are averaged over $10$ independent phase-lag realizations. The horizontal dashed lines in the right, center and left columns indicate the theoretical predictions for $1-r^\infty$ given by Eqs.~\eqref{eq:Theory16}, \eqref{eq:Theory17} and \eqref{eq:Theory20}, respectively.}
\label{fig1}
\end{figure*}

\subsection{Partially heterogeneous coupling and homogeneous coupling}\label{subsec:2.2}
Next we consider two special cases of coupling: partially heterogeneous coupling and homogeneous coupling. In the case of partial heterogeneity we assume that each oscillator has its own coupling function, i.e., $H_{ij}(\theta)=H_i(\theta)$. Carrying through an analysis similar to that outlined in Eqs.~(\ref{eq:Theory01})--(\ref{eq:Theory16}), we find that 
\begin{align}
1-r^\infty \simeq \frac{J(\bm{\hat{d}},L)}{2},\label{eq:Theory17}
\end{align}
where
\begin{align}
\hat{d}_i=H_i(0)d_i/H_i'(0).\label{eq:Theory19} 
\end{align}
Similar to the case of full heterogeneity, the contributions to the total erosion of synchronization from coupling frustration and network structure do not separate, this time due only to the construction of the new weighted degree vector $\bm{\hat{d}}$.

Finally, in the case of homogeneous coupling, each coupling function is identical, i.e., $H_{ij}(\theta)=H(\theta)$. This was the case studied in Ref.~\cite{Skardal2015PRE}, and the theory outlined in Eqs.~(\ref{eq:Theory01})--(\ref{eq:Theory16}) simplifies even more so, yielding
\begin{align}
1-r^\infty \simeq \frac{H^2(0)}{2H'^2(0)}J(\bm{d},L).\label{eq:Theory20}
\end{align}
Here, the right-hand side of Eq.~(\ref{eq:Theory20}) separates conveniently into the product of the square of the coupling frustration $h=|H(0)/\sqrt{2}H'(0)|$ and structural heterogeneity in the network, as measured by $J(\bm{d},L)$. This separation is a convenient property that is reminiscent of the separation of dynamics and structure in the analysis of synchronization of identical and nearly identical oscillators using the Master Stability Function approach~\cite{Pecora1998PRL,Sun2009EPL}.

For convenience, we summarize in Table~\ref{table} the expression for the total erosion of synchronization for each type of coupling we have considered, and indicate whether or not the contribution from coupling frustration and network structure are separable. We also note that in each of the three types of coupling we have considered here, we find that if no coupling frustration is present, i.e., all $H_{ij}(0)=0$, then the total erosion of synchronization is zero, i.e., $1-r^\infty=0$. This follows directly in the homogeneous case [Eq.~(\ref{eq:Theory20})] from the coefficient $H(0)$. In the fully and partially heterogeneous cases [Eqs.~(\ref{eq:Theory06}) and (\ref{eq:Theory07})] $H_{ij}(0)=0$ implies that $\bm{\tilde{d}}=\bm{0}$ and $\bm{\hat{d}}=\bm{0}$, respectively, and in turn $J(\bm{0},L)=0$ for any choice of $L$.

 \begin{table}[t!]
 \caption{\label{table}Summary of coupling types, total erosion of synchronization, and separability. Definition of the} synchrony alignment function $J$ is given in Eq.~\eqref{eq:Theory14}. 
\centering
\begin{tabular}{ l | c }
\hline
\hline
Coupling Type & $1-r^{\infty}$ \\
\hline
Hom., $H_{ij}(\cdot)\equiv H(\cdot)$& $\frac{H^2(0)}{2H'^2(0)}J(\bm{d},L)$ (separable) \\
Par. Het., $H_{ij}(\cdot)\equiv H_i(\cdot)$& $J(\bm{\hat{d}},L)/2$ (not separable) \\
Fully Het., $H_{ij}(\cdot)$ & $J(\bm{\tilde{d}},\tilde{L})/2$ (not separable) \\
\hline
\hline
\end{tabular}
\end{table}
~
\section{Numerics}\label{sec:3}
In this Section we illustrate the essential properties of the theory outlined above using several numerical examples. As a benchmark choice of coupling, from this point forward we consider Sakaguchi-Kuramoto-type coupling given by a sinusoid with an associated phase-lag, i.e., $H_{ij}(\theta)=\sin(\theta-\alpha_{ij})$~\cite{Sakaguchi1986PTP}. In particular, the parameter $\alpha_{ij}$ represents a phase-lag between oscillators $i$ and $j$, and can be taken from the range $\alpha_{ij}\in(-\pi/2,\pi/2)$ (although here we will restrict our focus on only non-negative values). In the case of full heterogeneity, each undirected link $(i,j)$ is then assigned its own phase-lag $\alpha_{ij}=\alpha_{ji}$, whereas in the case of partial heterogeneity each oscillator $i$ is assigned its own phase-lag $\alpha_i$. In the homogeneous case where a single phase-lag is chosen for all interactions, the coupling frustration is described by a single parameter $h=|H(0)/\sqrt{2}H'(0)|=|\tan(-\alpha)|/\sqrt{2}$.

For all simulations presented in the remainder of this paper, we solve Eq.~(\ref{eq:Kuramoto}) using a two-step Runge-Kutta method, a.k.a., Heun's method~\cite{Iserles2009}. We use a time step $\Delta t=4\times10^{-4}$, maintaining numerical stability for large coupling strengths by rescaling time by $K^{-1}$. Steady-state is obtained by discarding a significant transient of at least $10^5$ time steps. In each simulation we draw natural frequencies independently from the unit normal distribution. We will now present numerical results illustrating the effects of different phase-lags and network structures.

\begin{figure*}[t]
\centering
\includegraphics[width=0.85\linewidth]{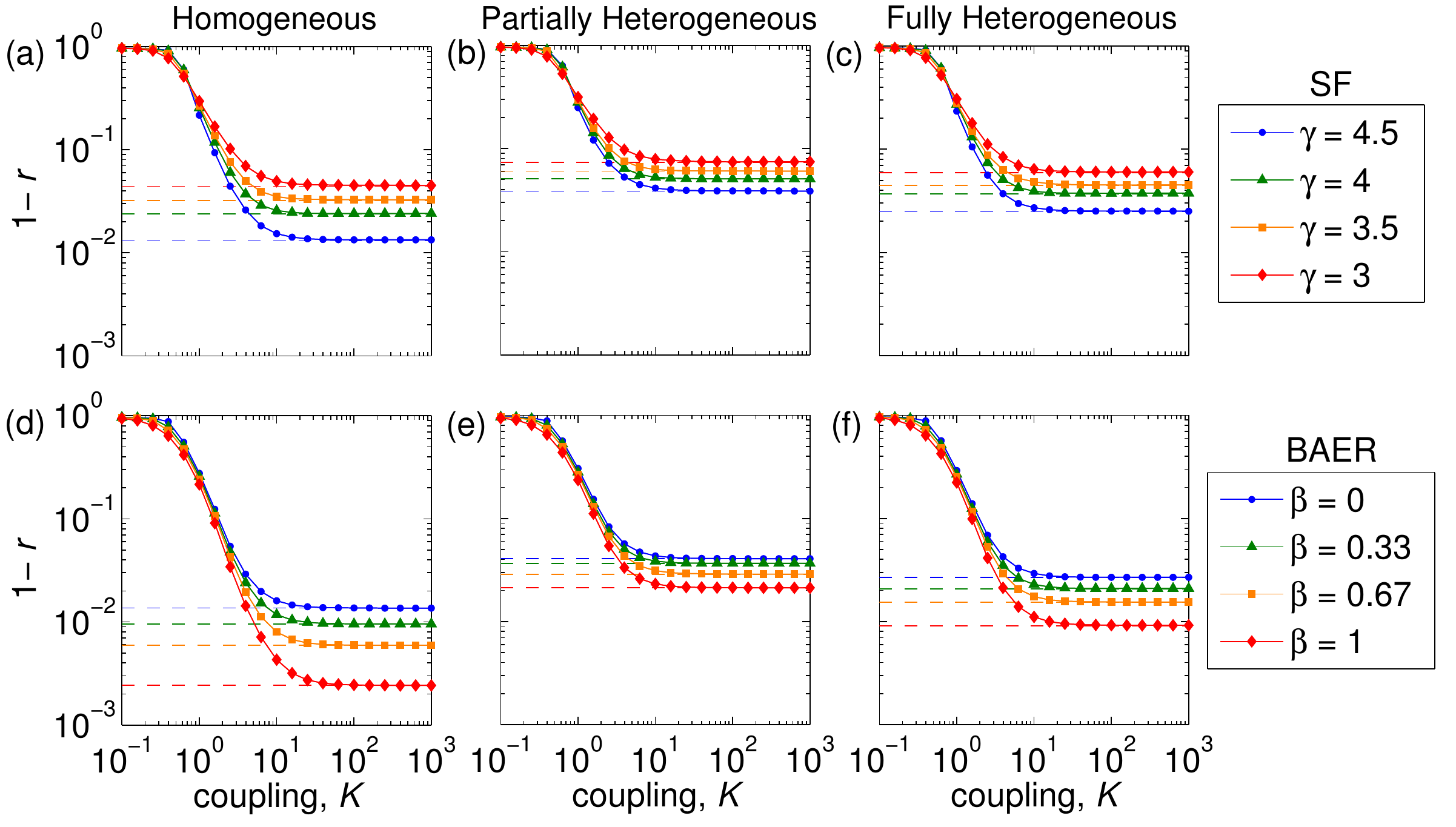}
\caption{(Color online) \textit{Erosion of synchronization: Effect of network structure.} Deviation from perfect synchronization $1-r$ vs coupling strength $K$ in various networks of size $N=1000$ and mean degree $\langle d\rangle=4$ using a fixed phase-lag parameter $\alpha=0.2$. Panels (a)--(c) display results for SF networks with homogeneous, partially heterogeneous, and fully heterogeneous coupling, respectively, and exponent $\gamma=4.5$ (blue circles), $4$ (green triangles), $3.5$ (orange squares), and $3$ (red diamonds). Panels (d)--(f) display results for BAER networks with homogeneous, partially heterogeneous, and fully heterogeneous coupling, respectively, and heterogeneity parameter $\beta=0$ (blue circles), $0.33$ (green triangles), $0.67$ (orange squares), and $1$ (red diamonds). Each data point represents an average over $10$ networks and $10$ independent phase-lag realizations for partially and fully heterogeneous couplings. The horizontal dashed lines in the right, center and left columns indicate the theoretical predictions for $1-r^\infty$ given by Eqs.~\eqref{eq:Theory16}, \eqref{eq:Theory17} and \eqref{eq:Theory20}, respectively.}
\label{fig2}
\end{figure*}

\subsection{Effect of phase-lags}\label{subsec:3.1}
We begin by considering the effect of different phase-lags on the erosion of synchronization for a given network. We first consider scale-free networks with power-law degree distribution $P(d)\propto d^{-\gamma}$ for $d\ge d_0$ built using the configuration model~\cite{Molloy1995}. In order to tune the mean degree $\langle d\rangle$ of each network, we set the minimum degree equal to $d_0=(\gamma-2)\langle d\rangle/(\gamma-1)$. We assign phase-lags according to a mean phase-lag parameter $\alpha$ as follows. In the case of homogeneous coupling, $\alpha$ describes the global phase-lag for the entire network. Otherwise, each distinct phase-lag ($\alpha_i$ or $\alpha_{ij}$) is drawn independently and uniformly from the interval $[0,2\alpha]$ to maintain a mean phase-lag of $\alpha$.

We begin our investigations by considering a SF networks size $N=1000$ with exponent $\gamma = 3$ and mean degree $\langle d\rangle=4$, and simulating the dynamics of Eq.~(\ref{eq:Kuramoto}) for homogeneous, partially heterogeneous, and fully heterogeneous coupling with several $\alpha$ values. We plot the resulting profiles $1-r$ vs $K$ for each type of coupling in Figs.~\ref{fig1}(a), (b), and (c), respectively, using a log-log scale. Results using phase-lag parameters $\alpha=0$, $0.05$, $0.1$, $0.15$, and $0.2$ are plotted using black circles, blue triangle, green squares, orange inverted triangles, and red diamonds, respectively. For the partially and fully heterogeneous cases, results represent an average over $10$ simulations with different phase-lag realizations. First, when no coupling frustration is present in the system (i.e., $\alpha=0$), the quantity $1-r$ decays as a power-law as $K$ is increased in each panel. This decay is observed to continue well past the windows shown. For nonzero values of $\alpha$, each curve $1-r$ initially decays, then saturates at a finite positive value, indicating that the perfectly synchronized state $r=1$ cannot be attained. For each of the three types of coupling heterogeneity, larger phase-lags yield a greater total erosion of synchronization. For each nonzero value of $\alpha$, we also plot the theoretical saturating value of $1-r^\infty$ in horizontal dashed lines [as given by Eqs.~\eqref{eq:Theory16}, \eqref{eq:Theory17} and \eqref{eq:Theory20}], which accurately predict the results from simulations. We also observe that the total erosion of synchronization tends to be larger on aggregate for partially heterogeneous coupling than for either fully heterogeneous or homogeneous coupling.

\begin{figure*}[t]
\centering
\includegraphics[width=0.85\linewidth]{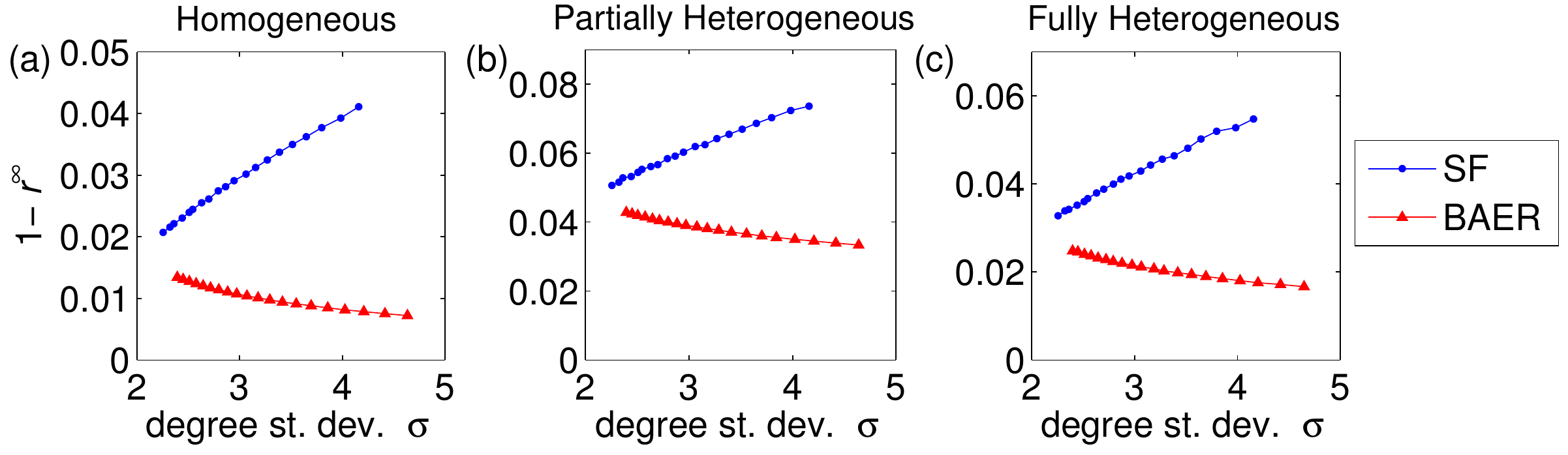}
\caption{(Color online) \textit{SF vs BAER network topologies.} Theoretically predicted total erosion of synchronization $1-r^\infty$ vs the standard deviation of the degree distribution $\sigma=\sqrt{\langle d^2\rangle-\langle d\rangle^2}$ for SF (blue circles) and $BAER$ (red triangles) networks of size $N=500$ with mean degree $\langle d\rangle=4$ using a phase-lag parameter $\alpha=0.2$. Panels (a)--(c) display results for homogeneous, partially heterogeneous, and fully heterogeneous couplings, respectively. Each data point represents an average over $1000$ network realizations and $10$ independent phase-lag realizations for partially and fully heterogeneous couplings.}
\label{fig3}
\end{figure*}

\subsection{Effect of network structure}\label{subsec:3.2}
Next we consider the effects of network structure on the erosion of synchronization. Using the same network model as in Sec.~\ref{subsec:3.1}, we first explore different topologies by varying the power-law exponent $\gamma$. In Figs.~\ref{fig2}(a), (b), and (c) we plot $1-r$ vs $K$ for networks of size $N=1000$ with mean degree $\langle d\rangle=4$ for homogeneous, partially heterogeneous, and fully heterogeneous coupling, respectively, for fixed $\alpha=0.2$. To vary the degree heterogeneity in each network, we use networks with corresponding exponent $\gamma=4.5$, $4$, $3.5$, and $3$, the results for which are plotted in blue circles, green triangles, orange squares, and red diamonds, respectively. Each curve represents the average over $10$ network realizations and phase-lag draws. For this network model we observe that networks with larger degree heterogeneity (smaller $\gamma$) yield larger total erosion of synchronization for all three types of coupling.

We contrast these results with results for another network model that allows for tunable degree heterogeneity. Specifically, we consider a network growth model~\cite{GomezGardenes2006PRE} that interpolates between Erd\H{o}s-R\'{e}nyi (ER) \cite{Erdos1960} type random networks and Barab\'{a}si-Albert (BA) \cite{Barabasi1999Science} preferential attachment type networks, which we summarize as follows. We prescribe a heterogeneity parameter $\beta\in[0,1]$ and a minimum degree $d_0$, and start with a fully-connected set of $d_0+1$ nodes. Next, nodes are added one-by-one to the network, each making $d_0$ links to the existing nodes. Each link can be made in one of two ways: with probability $\beta$ the link is made preferentially such that the new node connects to an existing node $i$ with probability $p_i\propto d_i-m$, and otherwise the link is made to a node uniformly at random. The parameter $m<d_0$ affects the final degree distribution. Nodes are added to the network until the desired size of the network $N$ is attained. If $\beta=0$, all links are constructed uniformly at random, and the resulting network has an ER-like topology. If $\beta=1$, the process is equivalent to a BA preferential attachment method, resulting in an SF structure with degree distribution $\gamma\approx 3-m/d_0$. Intermediate values of $\beta$ result in topologies that interpolate these two extreme cases. We refer to this model as the BAER model from this point forward. Importantly, the heterogeneity of the resulting networks can be tuned using the parameter $\beta$, with large (small) values yielding more (less) heterogeneous degree distributions.

We now explore the effect that tuning the heterogeneity parameter $\beta$ in the BAER model has on erosion of synchronization. In Fig.~\ref{fig2}(d), (e), and (f), we plot $1-r$ vs $K$ for BAER networks with homogeneous, partially heterogeneous, and fully heterogeneous coupling, respectively. Results are shown with fixed $\alpha=0.2$ for networks of size $N=1000$ with mean degree $\langle d\rangle=4$ ($d_0=2$) and $m=1.6$. We vary the heterogeneity parameter in each network, plotting the results from $\beta=0$, $0.33$, $0.67$, and $1$ in blue circles, green triangles, orange squares, and red diamonds, respectively. Each curve represents the average over $10$ network realizations and phase-lag draws. In contrast to the results obtained with SF networks (see the top row of Fig.~\ref{fig2} wherein increased heterogeneity yielded an increase in the total erosion of synchronization for all three coupling heterogeneities), we observe the opposite effect for BAER networks. In particular, as degree heterogeneity is increased (i.e., $\beta$ is increased) the total erosion of synchronization decreases for all three types of coupling heterogeneity. Finally, we note that similar to our observation made for Fig.~\ref{fig1}, in Fig.~\ref{fig2} one can also observe that the total erosion of synchronization tends to be larger in the case of partially heterogeneous coupling than for either fully heterogeneous or homogeneous coupling.

We next explore more closely the relationship between total erosion of synchronization and degree heterogeneity in the SF and BAER network models. Specifically, we compute for an ensemble of networks the predicted value $1-r^\infty$ and the standard deviation of the degree distribution $\sigma=\sqrt{\langle d^2\rangle-\langle d\rangle^2}$ for various $\gamma$ and $\beta$ and fixed $\langle d\rangle=4$ and $\alpha = 0.2$. In Fig.~\ref{fig3} we plot $1-r^\infty$ vs $\sigma$ for SF and BAER networks (blue circles and red triangles, respectively). Results are shown for homogeneous, partially heterogeneous, and fully heterogeneous coupling in panels (a), (b), and (c), respectively. Each data point represents the average over $1000$ networks of size $N=500$, each with $10$ independent phase-lag realizations for the cases with heterogeneous coupling. In all three panels, as $\sigma$ increases, the total erosion of synchronization increases for the SF model but decreases for the BAER model.

\section{Stability}\label{sec:4}
In this Section we shift our focus to the stability of the stationary state given in Eq.~(\ref{eq:Theory10}). We consider here the limit of strong coupling strength, i.e., $K\to\infty$. The linear stability of this state is dictated by the spectrum of the scaled Jacobian matrix $\widetilde{DF}=DF/K$, whose entries are given by
\begin{align}
\widetilde{DF}_{ij} = \left\{\begin{array}{rl} A_{ij}H_{ij}'(\theta_j^*-\theta_i^*) &\text{if } i\ne j \\ -\sum_{j\ne i}A_{ij}H_{ij}'(\theta_j^*-\theta_i^*) & \text{if }i = j. \end{array}\right.\label{eq:Theory22}
\end{align}
Like the Laplacian matrix, each row of $\widetilde{DF}$ sums to zero, yielding a trivial zero eigenvalue that corresponds to the translational invariance of the dynamics in Eq.~(\ref{eq:Kuramoto}). A stationary solution $\bm{\theta}^*$ is then linearly stable if the real parts of all nontrivial eigenvalues are negative, or conversely, if no eigenvalue has positive real part. More precisely, given the eigenvalues $\lambda_i$ of $\widetilde{DF}$ we define $\lambda^{\textrm{max}}_{\text{re}} = \max_i \{\textrm{Re}(\lambda_i): \lambda_i\not=0\} $, such that $\lambda^{\textrm{max}}_{\text{re}} <0$ ($>0$) implies that the state $\bm{\theta^*}$ is linearly stable (unstable).

To guide our analysis, we first consider the case of no coupling frustration, $H_{ij}(0)=0$, for which $\bm{\theta}^*=\bm{0}$. This follows from the fact that each entry of the vector $\bm{\tilde{d}}$ is zero and implies that the scaled Jacobian is negatively proportional to the unweighted Laplacian, $\widetilde{DF}\propto-L$. Assuming that the network is strongly connected, the nontrivial eigenvalues of $L$ all have positive real part, implying that the nontrivial eigenvalues of $\widetilde{DF}$ have all negative real part, and thus the solution is stable. 

Next, we consider solutions under the presence of coupling frustration. Given that there exists a spectral gap for the unweighted Laplacian, a sufficiently small amount of coupling frustration can be found such that the solution $\bm{\theta}^*$ is guaranteed to remain stable. The extent to which coupling frustrations can be increased without destabilizing the solution, however, is unclear. To shed some light on this, we consider the structure of $\widetilde{DF}$. For the typical case of $H_{ij}(0)>0$ and small enough $|\theta_j^*-\theta_i^*|$, the scaled Jacobian consists of negative diagonal entries with non-negative off-diagonal entries. In particular, as long as this structure is maintained, $\widetilde{DF}$ remains negative semi-definite and its spectrum remains bounded in the left-half complex plane. (This can be shown, for instance, by using the {\it Gershgorin circle theorem}~\cite{Golub}.) A necessary condition for instability is then given by at least one off-diagonal entry becoming negative, which allows for the possibility of one or more eigenvalues crossing the imaginary axis. In the case of homogeneous coupling, we showed in Ref.~\cite{Skardal2015PRE} that a necessary condition for the loss of linear stability is given by 
\begin{align}
\min_{A_{ij\ne0}}H'\left(\frac{H(0)}{H'(0)}\left([L^\dagger\bm{d}]_j-[L^\dagger\bm{d}]_i\right)\right)<0.\label{eq:Theory23}
\end{align}
For partially and fully heterogeneous coupling, this condition becomes, respectively,
\begin{align}
\min_{A_{ij\ne0}}H_{i}'\left([L^\dagger\bm{\hat{d}}]_j-[L^\dagger\bm{\hat{d}}]_i\right)<0,\label{eq:Theory24}
\end{align}
and
\begin{align}
\min_{A_{ij\ne0}}H_{ij}'\left([\tilde{L}^\dagger\bm{\tilde{d}}]_j-[\tilde{L}^\dagger\bm{\tilde{d}}]_i\right)<0.\label{eq:Theory25}
\end{align}
When the appropriate equation of Eqs.~(\ref{eq:Theory23})--(\ref{eq:Theory25}) is false the synchronized state is guaranteed to be linear stable, but if it holds true, this admits the possibility that it is unstable.

\begin{figure}[t]
\centering
\includegraphics[width=0.90\linewidth]{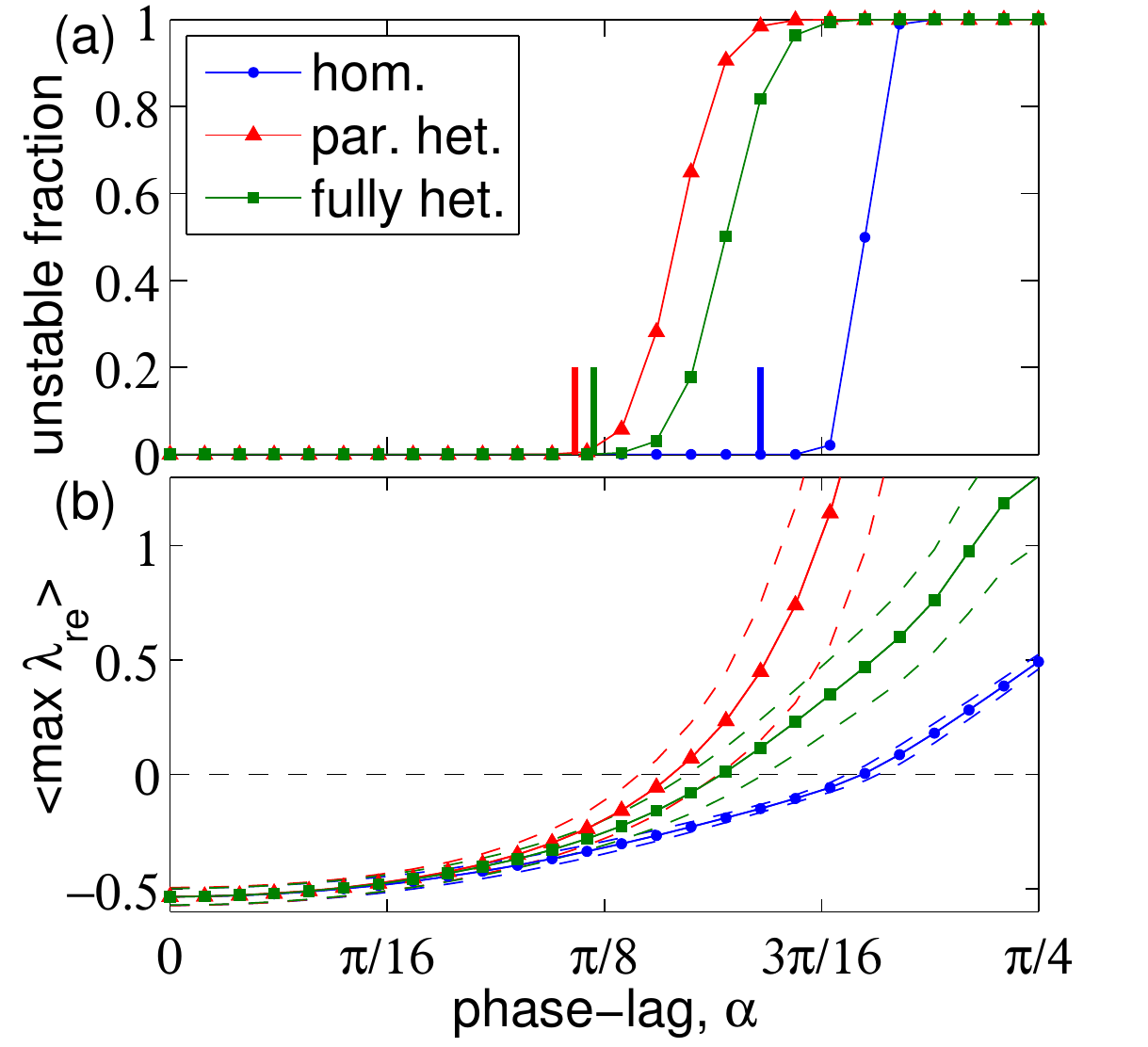}
\caption{(Color online) \textit{Stability of the synchronized solution.} For BAER networks of size $N=500$ and $\beta = 0$, (a) the fraction of solutions found to be unstable for homogeneous (blue circles), partially heterogeneous (red triangles), and fully heterogeneous (green squares) coupling. Thick vertical markers indicate the critical value $\alpha$ where the necessary instability conditions given by Eqs.~(\ref{eq:Theory10})--(\ref{eq:Theory10c}). All results indicate mean values computed across 100 simulations per each $\alpha$ value. (b) The mean real part of the largest nontrivial eigenvalue $\langle\max\lambda_{\text{re}}\rangle$ (symbols) plus and minus (dashed curves) the standard deviation. Results represent the average over $100$ network realizations and $10$ independent phase-lag realizations for partially and fully heterogeneous coupling.}
\label{fig4}
\end{figure}

To illustrate the utility of Eqs.~(\ref{eq:Theory23})--(\ref{eq:Theory25}), we study the stability BAER networks of size $N=500$ with $\beta=0$ with homogeneous, partially heterogeneous, and fully heterogeneous coupling. (We note that results obtained using other parameters and network models were found to be similar.) First, using a sample of $1000$ networks each with $10$ independent realizations of phase-lags for each value of $\alpha$, we plot in Fig.~\ref{fig4}(a) the fraction of instances where the synchronized state was found to be unstable. Results for homogeneous, partially heterogeneous, and fully heterogeneous coupling are plotted using blue circles, red triangles, and green squares, respectively. We also calculate the critical value of $\alpha$ for which the conditions given in Eqs.~(\ref{eq:Theory23})--(\ref{eq:Theory25}) first become true on average, which we denote with using thick vertical markers. Our results indicate that these conditions do in fact provide a reasonably good lower bound for the transition to instability.

In Fig.~{\ref{fig4}}(b) we plot the average real part of the largest nontrivial eigenvalue $\lambda^{\textrm{max}}_{\text{re}} $ of $\widetilde{DF}$ calculated from these simulations, denoting the standard deviations using dashed curves. Recall that for a given system, $\lambda^{\textrm{max}}_{\text{re}} <0$ and $\lambda^{\textrm{max}}_{\text{re}} >0$ imply stability and instability, respectively. Therefore, the transitions from stability to instability shown in panel (a) are driven by the dependence of $\lambda^{\textrm{max}}_{\text{re}} $ on $\alpha$. Naturally, for small $\alpha$ the values of $\lambda^{\textrm{max}}_{\text{re}} $ do not depend strongly on the type of coupling heterogeneity. However, as $\alpha$ increases, the dependence of $\lambda^{\textrm{max}}_{\text{re}}$ on $\alpha$ is varies for each type of coupling heterogeneity, which leads to the different transitions shown in panel (a). For example, $\lambda^{\textrm{max}}_{\text{re}} $ grows faster for the partially and fully heterogeneous couplings than for homogeneous coupling, which leads to a transition from stability to instability that occurs for smaller $\alpha$. Moreover, the standard deviation in $\lambda^{\textrm{max}}_{\text{re}}$ is much smaller for homogeneous coupling than for the other couplings, and thus the transition from stability to instability is much more abrupt for homogeneous coupling than for the heterogeneous couplings.

\section{Special network structures}\label{sec:5}
In all of the results presented above, both for predicting the total erosion of synchronization and classifying the stability of the solution, a key ingredient is the (weighted or unweighted) network Laplacian, and especially its spectral properties. While the spectral properties for realizations of certain networks models can vary significantly~\cite{Carlson2011Chaos}, for a handful of special network structures the spectral properties are well-known. Such network structures, such as stars and chains, are used in various studies as either small motifs or modules~\cite{Milo2002Science,Taylor2011PRE}, collections of which make up a full network, or on their own, serving as examples for various dynamical phenomena~\cite{GomezGardenes2011PRL,Leyva2012PRL,Bergner2012PRE}. In order to obtain analytical predictions for the total erosion of synchronization in such networks, we focus here on the case of homogeneous coupling where the unweighted Laplacian and degree vector can be used. Furthermore, while we have observed previously that results for homogeneous coupling differ from heterogeneous coupling quantitatively, we note that homogeneous coupling does in fact serve as a predictive benchmark, displaying qualitatively similar results to heterogeneous coupling.

We will begin by presenting analytical results for the star and chain networks, where the total erosion of synchronization can be predicted in terms of the network size. We note that all networks with regular structure, i.e., all nodes having the same degree, such as the ring or periodic lattices, trivially yield zero total erosion of synchronization, and therefore we forgo any more consideration of these networks here. Finally, we provide analytical approximations for the total erosion of synchronization in Watts-Strogatz (WS) networks~\cite{Watts1998Nature}, an important network model where progress is currently being made for the analytical approximation of spectral properties~\cite{Grabow2012PRL}.

\begin{figure}[t]
\centering
\includegraphics[width=0.90\linewidth]{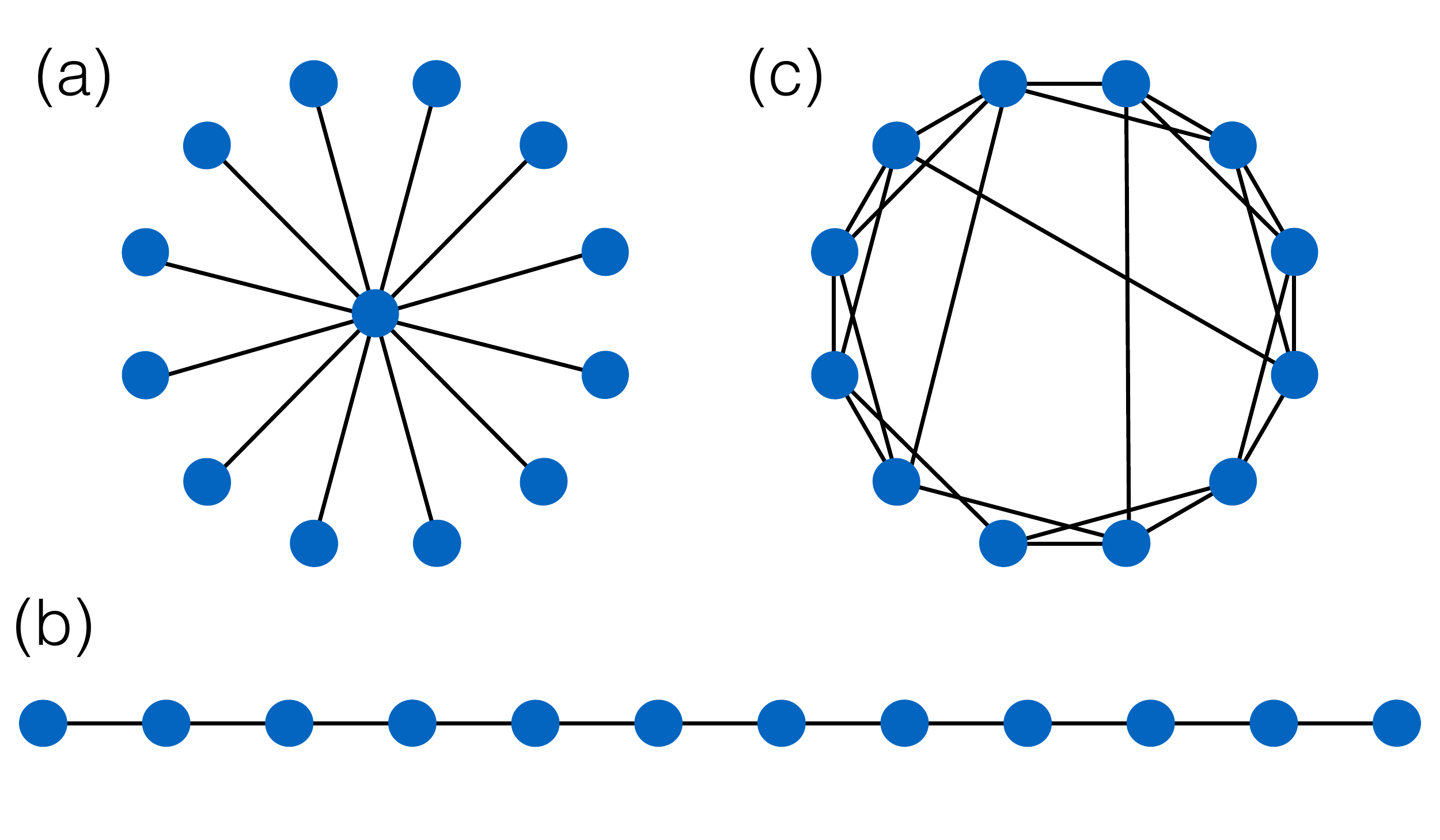}
\caption{(Color online) \textit{Special network structures.} Example illustrations of (a) star, (b) chain, and (c) Watts-Strogatz networks.}
\label{fig5}
\end{figure}

\subsection{Star network}\label{subsec:5.1}
We begin by considering the star network--a network consisting of a single ``hub'' node connected to many ``leaves'' [see Fig.~\ref{fig5}(a)]. For a star network of size $N$, we index the hub $i=1$ and the leaves $i=2,\dots,N$. The network Laplacian matrix is then given by
\begin{align}
L = \begin{bmatrix}
N-1 & -1 & -1 & \ldots & -1 \\
-1 & 1 & 0 & \ldots & 0 \\
-1 & 0 & 1 & \ldots & 0 \\
\vdots & \vdots & \vdots & \ddots & \vdots \\
-1 & 0 & 0 & \ldots & 1
\end{bmatrix},\label{eq:Star01}
\end{align}
and the degree vector is given by $\bm{d} = [N-1,1,\dots,1]^T$. It is easy to show that the eigenvalues of $L$ are $\lambda_1=0$, $\lambda_2=\dots=\lambda_{N-1}=1$, and $\lambda_N=N$. The normalized eigenvectors corresponding to $\lambda_1$ and $\lambda_N$ are $\bm{v}^1=[1,\dots,1]^T/\sqrt{N}$ and $\bm{v}^N=[N-1,-1,\dots,-1]^T/\sqrt{N^2-N}$ and the degree vector can be written as the linear combination
\begin{align}
\bm{d} = \frac{2N-2}{\sqrt{N}}\bm{v}^1 + (N-2)\sqrt{1-N^{-1}}\bm{v}^N.\label{eq:Star02}
\end{align}
Finally, when evaluating $J(\bm{d},L)$, only the $\bm{v}^N$ part contributes, and after simplification we obtain
\begin{align}
J(\bm{d},L) =\frac{(N-2)^2(1-N^{-1})}{N^3}.\label{eq:Star03}
\end{align}

Interestingly, the behavior of $J(\bm{d},L)$ for the star network is not monotonic with the network size $N$. At $N=2$, $J(\bm{d},L)$ is zero, corresponding to fact that the degree vector is constant, immediately above which $J(\bm{d},L)$ increases. A maximum is reached at $N=4+2\sqrt{2}$, after which $J(\bm{d},L)$ decreases for all larger $N$, scaling like $J(\bm{d},L)\sim N^{-1}$. Restricting $N$ to integers greater than or equal to 2, this implies that the maximum total erosion of synchronization for homogeneous coupling in a star network is obtained at a size of $N=7$ (i.e., six leaf nodes), above and below which the total erosion is less. 

\begin{figure}[t]
\centering
\includegraphics[width=0.90\linewidth]{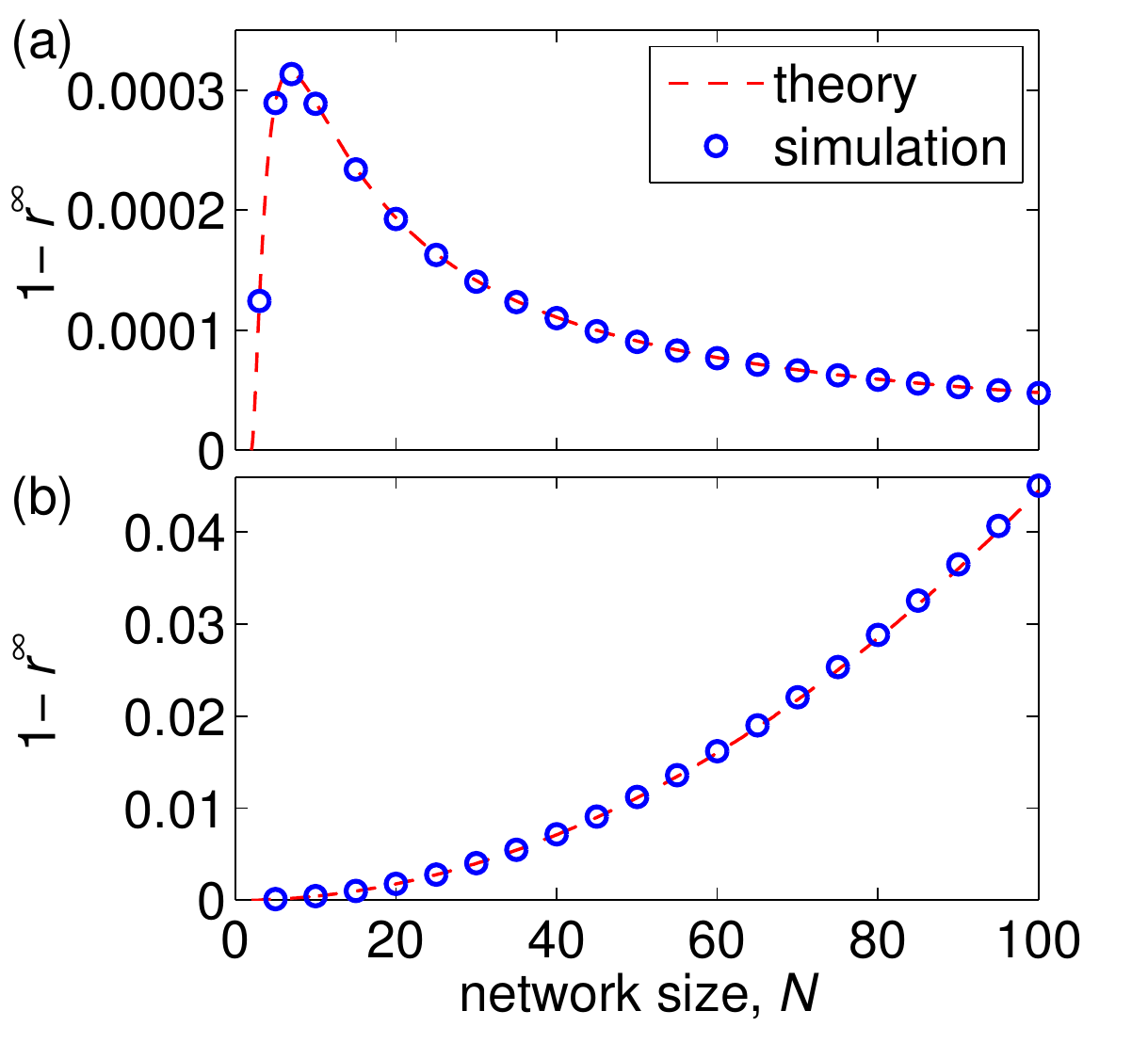}
\caption{(Color online) \textit{Erosion of synchronization: stars and chains.} The total erosion of synchronization $1-r^\infty$ as a function of network size $N$ for (a) star and (b) chain networks using $\alpha=0.1$ and $0.04$, respectively. Theoretical results given by Eq.~\eqref{eq:Theory20} and either (a)~Eq.~\eqref{eq:Star03} or (b) Eq.~\eqref{eq:Chain03} are plotted as dashed red curves, and results from simulations are plotted with blue circles.}
\label{fig6}
\end{figure}

In Fig.~\ref{fig6}(a) we plot our theory vs simulations using star networks of various sizes and setting $\alpha=0.1$. The theoretical curve is obtained using Eq.~(\ref{eq:Star03}) with Eq.~(\ref{eq:Theory20}) and plotted as the dashed red curve, and results from simulations are plotted using blue circles. We note excellent agreement between theory and simulation, both of which capture the non-monotonicity of $1-r^\infty$ depending on network size.

\subsection{Chain network}\label{subsec:5.2}
Next, we consider the undirected chain--a network consisting of sequentially linked nodes with end nodes indexed $i=1$ and $N$ [see Fig.~\ref{fig5}(b)]. The Laplacian of the chain is given by,
\begin{align}
L = \begin{bmatrix}
1 & -1 & \ldots & 0 & 0 \\
-1 & 2 & \ldots & 0 & 0 \\
\vdots & \vdots & \ddots & \vdots & \vdots \\
0 & 0 & \ldots & 2 & -1 \\
0 & 0 & \ldots & -1 & 1
\end{bmatrix},\label{eq:Chain01}
\end{align}
and the degree vector is given by $\bm{d}=[1,2,\dots,2,1]^T$. It is straight forward to show that the eigenvalues of $L$ are given by $\lambda_j=4\sin^4[\pi(j-1)/2N]$, and the corresponding eigenvectors are $\bm{v}^1=[1,\dots,1]^T/\sqrt{N}$ and $\bm{v}^j$ with entries given by $v_i^j=\sqrt{2/N}\cos[\pi(j-1)(2i-1)/2N]$ for $j\ge2$. The degree vector can thus be written as
\begin{align}
\bm{d} = 2\sqrt{N}\bm{v}^1-\bm{e}^1-\bm{e}^N,\label{eq:Chain02}
\end{align}
where $\bm{e}^j$ is the canonical basis vector with entries $e_i^j=\delta_{ij}$. Again, the eigenvector $\bm{v}^1$ does not contribute to $J(\bm{d},L)$, and after some algebra, we obtain
\begin{align}
J(\bm{d},L) &= \frac{1}{N^2}\sum_{j=2}^N\frac{\left(\cos\varphi_j+\cos\left[\varphi_j(2N-1)\right]\right)^2}{8\sin^4\varphi_j},\label{eq:Chain03}
\end{align}
where $\varphi_j=\pi(j-1)/2N$.

In Fig.~\ref{fig6}(a) we study the total erosion of synchronization for chains of various sizes using $\alpha=0.04$. The dashed red curve indicates our theory, which is obtained using Eq.~(\ref{eq:Chain03}) with Eq.~(\ref{eq:Theory20}), and results from simulations are plotted using blue circles. These are in excellent agreement. We find that unlike star networks, the total erosion of synchronization for chains increases monotonically with increasing size $N$. The increasing behavior of $J(\bm{d},L)$ for the chain is not only surprising in light of the results from the star network, but also in comparison to the ring. In particular, rings and chains of the same size differ by only the addition/subtraction of a single undirected link, and they in fact share similar spectral properties. However, whereas the total erosion of synchronization is trivially zero in the ring since the degree vector is constant, the chain yields a large total erosion of synchronization that increases with size. It is also worth noting that on the chain, the dynamics take a long time to relax to steady-state, which can be explained by the fact that the time-scale for relaxation $\tau$ is dictated by the inverse of the smallest non-trivial eigenvalue $s_2=4\sin^4\left(\pi/2N\right)$, which approaches zeros as $N$ approaches infinity.

\subsection{Watts-Strogatz networks}\label{subsec:5.3}
Finally, we consider the Watts-Strogatz (WS) network model~\cite{Watts1998Nature}. This popular model provides an interpolation between regular periodic lattice-type networks and random networks which display the small-world property. Here we consider the simple case of ring-like WS networks [see Fig.~\ref{fig5}(c)]. Given a set of $N$ nodes arranged in a ring and a chosen uniform degree $d$ (assuming to be even), each node is connected to the $d/2$ closest neighbors on each side. Next, given a rewiring probability $q$, each link is either rewired or not rewired, with probabilities $q$ and $1-q$, respectively. Each rewired link is then replaced with a link connecting one of the original nodes (chosen randomly) and another node that is chosen uniformly at random from the remaining nodes. Thus, for extreme values $q=0$ and $q=1$, the resulting network is a perfect ring or an Erd\H{o}s-R\'{e}nyi type network~\cite{Erdos1960}.

\begin{figure}[t]
\centering
\includegraphics[width=0.90\linewidth]{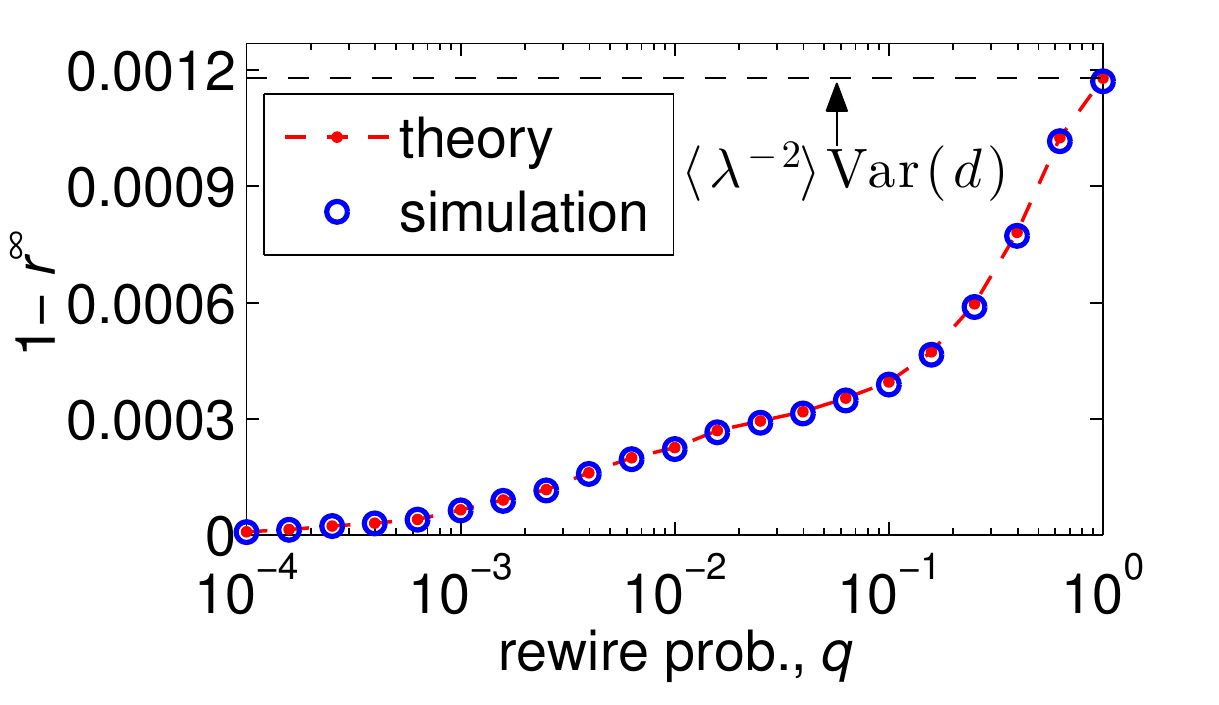}
\caption{(Color online) \textit{Erosion of synchronization: Watts-Strogatz networks.} The total erosion of synchronization $1-r^\infty$ as a function of the rewire probability $q$ for WS networks of size $N=500$ with mean degree $\langle d\rangle=40$ using a homogeneous phase-lag $\alpha=0.4$. Theory given by Eq.~\ref{eq:Theory22} and simulations of the dynamics are plotted by the red dashed red curve and blue circles, respectively. We also plot Eq.~\eqref{eq:WS01} by the horizontal dashed line, which is in excellent agreement when $q=1$.}\label{fig7}
\end{figure}

While precise expressions for the total erosion of synchronization in general WS networks are difficult to obtain, they can be found for the limiting values of $q=0$ and $1$. For $q=0$ the network is a regular one-dimensional lattice, and thus $J(\bm{d},L)=0$ trivially since the degree vector $\bm{d}$ is constant. For $q=1$ on the other hand, $J(\bm{d},L)$ can be approximated based a simplifying assumption. In particular, given the completely random network structure, we assume that the eigenvalues $\lambda_j$ and the projections of the eigenvectors onto the degree vector $\langle\bm{v}^j,\bm{d}\rangle$ are uncorrelated, allowing us to separate the synchrony alignment function:
\begin{align}
J(\bm{d},L) &= \frac{1}{N}\sum_{j=2}^N\frac{\langle\bm{v}^j,\bm{d}\rangle^2}{\lambda_j^2}\nonumber\\
 &\approx \left(\frac{1}{N}\sum_{j=2}^N\lambda_j^{-2}\right)\left(\sum_{j=2}^N\langle\bm{v}^j,\bm{d}\rangle^2\right)\nonumber\\
 &=\langle \lambda^{-2}\rangle\text{Var}(d).\label{eq:WS01}
\end{align}
where we have used that the second term in the product is equal to the variance of the degree distribution. In principle, Eq.~(\ref{eq:WS01}) can be used to approximate $J(\bm{d},L)$ directly from a given network's properties (i.e., the degree vector and eigenvalues), or rom using known asymptotic results for the class of networks. For example, it is well known for ER networks that the degree distribution is binomial and the eigenvalue distribution is in certain cases well-approximated based on a semi-circle law.

We test this approximation and investigate the behavior of WS networks for intermediate rewiring probabilities by comparing the predicted total erosion of synchronization given by Eq.~\eqref{eq:Theory20} and the results from simulations for several WS networks. In Fig.~\ref{fig7} we plot the total erosion of synchronization $1-r^\infty$ vs the rewire probability $q$ for WS networks of size $N=500$ with mean degree $\langle d\rangle=40$ using $\alpha=0.4$. Theoretical predictions are plotted by the dashed red curve, and results from simulations are plotted in blue circles. Each data point represent an average over $20$ network realizations. In addition, we plot by the horizontal dashed line our approximation for $q=1$ given by Eq.~\eqref{eq:WS01}, which is in excellent agreement with both our simulations and Eq.~\eqref{eq:Theory20} when $q=1$.

\section{Discussion}\label{sec:6}
In this Article we have studied erosion of synchronization in networks of coupled oscillators, whereby perfect synchronization is unattainable even in the limit of infinite coupling strength and is a phenomenon that arises in the presence of both coupling frustration and structural heterogeneity. We have generalized previous results to the important case of heterogeneous coupling, allowing for the interactions between different pairs of network neighbors to be described by different functions. As compared to homogeneous coupling, where a single coupling function describes all of the interaction in the network, the theoretical predictions for heterogeneous coupling become more complicated. While the theoretical prediction for the total erosion of synchronization separates into the product of terms describing the coupling frustration and network structure in the homogeneous case [see Eq.~\eqref{eq:Theory20}], it does not when coupling is heterogeneous [see Eqs.~\eqref{eq:Theory16} and \eqref{eq:Theory17}]. However, our predictions show that the presence of heterogeneity in coupling frustrations amplifies the total erosion of synchronization; even when the mean frustration is the same as the homogeneous case, heterogeneity in the frustration increases the deviation from the perfectly synchronized state, as measured by $1-r$. Additionally, the heterogeneity in the coupling functions reduces the range of coupling frustrations for which the synchronized solution remains stable [see Fig.~\ref{fig4}]. 

We have also studied erosion of synchronization in some special network structures. In the case of homogeneous coupling, we have derived analytical results for the star and chain networks. Remarkably, in the star network the total erosion of synchronization is maximized for a star of size $N=7$ (i.e., $6$ leaves) and decays when $N$ is either increased and decreased. In particular, for large $N$ we find that $1-r^\infty$ decays as $N^{-1}$. In the case of the chain network, the total erosion of synchronization increases as the chain is lengthened---that is, $1-r^\infty$ increases with $N$ for $N\leq100$ (and it will have to decrease at some point because $r^\infty\geq0$). Finally, we investigated the case of Watts-Strogatz networks and provided further analytical results for the limiting cases in which the rewiring probability is $q=0$ or $1$.

\section*{Acknowledgements}
This work was funded in part by the James S. McDonnell Foundation (PSS and AA), NSF Grant No. DMS-1127914 through the Statistical and Applied Mathematical Sciences Institute (DT), Simons Foundation Grant No. 318812 (JS), Spanish DGICYT Grant No. FIS2012-38266 (AA), and FwET Project No. MULTIPLEX (317532) (AA).

\bibliographystyle{plain}

\end{document}